\documentclass[10pt]{revtex4}
\usepackage[dvips]{graphicx}
\usepackage{amsthm,latexsym,amssymb,amsfonts,array}

\begin{document}

\title{Born-Infeld electrostatics in the complex plane}
\author{Rafael Ferraro \thanks{Member of Carrera del
Investigador Cient\'{\i}fico (CONICET, Argentina)}\\ Instituto de
Astronom\'\i a y F\'\i sica del Espacio, Casilla de Correo 67,
Sucursal 28, 1428 Buenos Aires, Argentina\\ Departamento de F\'\i
sica, Facultad de Ciencias Exactas y Naturales, Universidad de
Buenos Aires, Ciudad Universitaria, Pabell\'on I, 1428 Buenos
Aires, Argentina}


\begin{abstract}
The complex method to obtain 2-dimensional Born-Infeld electrostatic
solutions is presented in a renewed form. The solutions are
generated by a holomorphic seed that makes contact with the
Coulombian complex potential. The procedure is exemplified by
solving the Born-Infeld multipolar configurations. Besides, it is
shown that the attractive force between two equal but opposite
charges is lower than its Coulombian partner; it decreases up to
vanish when the charges approach each other below a distance ruled
by the Born-Infeld constant.
\end{abstract}



\maketitle

\section{Introduction}
Born-Infeld electrodynamics was born as a non-linear extension of
Maxwell's equations able to render finite the self-energy of the
point-like charge \cite{B}-\cite{BI4}. After decades of relative
oblivion, Born-Infeld theory regained a prominent place in
theoretical physics because of its role in the low energy dynamics
of strings and branes \cite{Frad}-\cite{Tsey2}. Born-Infeld theory
is distinguished as the only extension of Maxwell's theory having
causal propagation \cite{Pleb,Deser} and absence of birefringence
\cite{Boillat,Novello}. While its free plane waves do not differ
from Maxwell's ones, the Born-Infeld non-linearity provides
interactions among plane waves \cite{barba,ferr} or between plane
waves and static fields
\cite{Pleb,Boillat,5pleb,Aiello,ferr2,gibbons} that substantially
change the physics of propagation. Born-Infeld electrodynamics
possesses a magnitude $b$ with units of field that rules the scale
of field at which Maxwell's theory is recovered (in the same way
that $c$ rules the Newtonian limit of relativistic mechanics). In
this paper we will continue the program to obtain Born-Infeld
electrostatic solutions in the Euclidean plane. This program began
very early with the articles by Pryce \cite{Pryce1,Pryce2}, who used
the complex analysis to establish the main features of the
electrostatic configurations for isolated point-like charges. In
Sections 2 and 3 we will present the complex method to generate
2-dimensional electrostatic solutions in a renewed and cleaner way.
Recently, the multipolar configurations were worked out \cite{Fer};
these solutions displayed some physically undesirable features that
will be healed in Sections 5 and 6. Particular features of the
dipole field are examined in Sections 7-9, together with general
complex expressions for Born-Infeld electrostatic forces and
energies. Section 10 describes the solution for two separated equal
but opposite charges. It is shown that the attractive force reaches
a maximum value at a non-null distance, and then it decreases up to
vanish when the charges meet together. Some important
characteristics of the holomorphic functions that generate
Born-Infeld solutions for point-like charges are discussed in
Sections 10 and 11.

\section{Born-Infeld theory}
Like Maxwell's theory, vacuum Born-Infeld electrodynamics is
summarized in two equations:
\begin{equation}
dF\, =\, 0\ ,  \label{eq0}
\end{equation}
\begin{equation}
d\ast \mathcal{F}\, =\, 0\ .  \label{eq1}
\end{equation}
The 2-form $F$ is the electromagnetic field, and $\mathcal{F}$ is
the 2-form
\begin{equation}
\mathcal{F}\ \equiv \ \frac{F-\frac{P}{b^{2}}\ \ast
F}{\sqrt{1+\frac{2S}{b^{2}}-\frac{P^{2}}{b^{4}}}}\ ,  \label{eq2}
\end{equation}
where $S$\ and $P$\ are the scalar and pseudoscalar field
invariants,
\begin{equation}
S\ =\ \frac{1}{4}\ F_{\mu \nu }F^{\mu \nu }\ =\ \frac{1}{2}\
(|\mathbf{B}|^{2}-|\mathbf{E}|^{2})\ ,  \label{eq3}
\end{equation}
\begin{equation}
P\ =\ \frac{1}{4}\ ^{\ast }F_{\mu \nu }F^{\mu \nu }\ =\
\mathbf{E}\cdot \mathbf{B}\ . \label{eq4}
\end{equation}
and $\ast $ is the Hodge star operator. The $^{\ast }F_{\mu \nu }$'s
--the components of $\ast F$-- compose the dual field tensor, i.e.
the tensor resulting from exchanging the roles of the electric and
magnetic fields: $\mathbf{E}\longleftrightarrow {-\mathbf{B}}$.
Born-Infeld equation (\ref{eq0}) does not differ from those
Maxwell's equations governing the curl of $\mathbf{E}$ and the
divergence of $\mathbf{B}$. It allows to write the field as the
exterior derivative of a 1-form $A$ (the electromagnetic potential):
$F=dA$ (i.e. $F_{\mu \nu }=\partial _{\mu }A_{\nu }-\partial _{\nu
}A_{\mu }$). Instead, Born-Infeld equation (\ref{eq1}) --which means
$\partial_{\nu}(\sqrt{-g}\, \mathcal{F}^{\mu\nu })=0$-- departs from
the respective Maxwell's one; however Maxwell's equation $d\ast F=0$
is recovered in the limit $b\rightarrow \infty $. Eq.~(\ref{eq1})
can be derived from the Born-Infeld scalar Lagrangian:
\begin{equation}
L[A]=\frac{b^{2}}{4\,\pi }\;\left(
1-\sqrt{1+\frac{2S}{b^{2}}-\frac{P^{2}}{b^{4}}}\right)\ ,
\label{eq5}
\end{equation}
which goes to the Maxwell Lagrangian $L[A]=-S/(4\pi)$ when
$b\rightarrow\infty$. Those solutions having $S=0=P$ (``free waves")
are shared by both Maxwell and Born-Infeld theories. The
energy-momentum tensor is (for metric signature $+---$)
\begin{equation}
T_{\,\mu \,\nu }=\frac{2}{\sqrt{-g}}\,\frac{\partial
(\sqrt{-g}\,L)}{\partial g^{\mu \nu }}=-\frac{1}{4\pi }\,F_{\mu
\,\rho }\,\mathcal{F}_{\nu }^{\;\,\rho }-\frac{b^{2}}{4\pi }\ g_{\mu
\,\nu }\ \left( 1-\sqrt{1+\frac{2S
}{b^{2}}-\frac{P^{2}}{b^{4}}}\right)\ . \label{temunu}
\end{equation}

\section{Born-Infeld electrostatics in 2 dimensions}
For electrostatic configurations, Eqs.~(\ref{eq0}, \ref{eq1}) reduce
to
\begin{equation}
\mathbf{\nabla }\times \mathbf{E}\, =\, 0\ ,  \label{eq6}
\end{equation}
\begin{equation}
\mathbf{\nabla }\cdot \mathbf{D}\, =\, 0\ ,  \label{eq7}
\end{equation}
where%
\begin{equation}
\mathbf{D}\equiv
\frac{\mathbf{E}}{\sqrt{1-\frac{|\mathbf{E}|^{2}}{b^{2}}}}\ .
\label{eq8}
\end{equation}
In the Euclidean plane $(x,y)$ the vector language can be rephrased
in the language of complex differential forms. Any function $f(x,y)$
can be written as $f(z,\overline{z})$, since $x=(z+\overline{z})/2$,
$y=-i(z-\overline{z})/2$. Thus the 1-form
\begin{equation}
F=E_{x}\ dx+E_{y}\ dy\ , \label{campo1}
\end{equation}
can be rewritten as
\begin{equation}
F=\frac{1}{2}\ (E_{x}-iE_{y})\ dz+\frac{1}{2}\ (E_{x}+iE_{y})\
d\overline{z}\ . \label{campo2}
\end{equation}
The 1-form (\ref{campo1}, \ref{campo2}) is the electric field (the
original 2-form $F$ of Eq.~(\ref{eq0}) has become a 1-form once the
$t$ coordinate has been suppressed in the static approach). We will
call $E$ the complex function
\begin{equation}
E(z,\overline{z})\equiv E_{x}-iE_{y}\ .  \label{campo3}
\end{equation}
Analogously, it is
\begin{equation}
\mathcal{F}=\frac{1}{2}\ (D_{x}-iD_{y})\ dz+\frac{1}{2}\
(D_{x}+iD_{y})\ d\overline{z}\ ,  \label{campo4}
\end{equation}
and $D\equiv D_{x}-iD_{y}$. The curl and the divergence in 2
dimensions can be retrieved from the operator $\partial /\partial
\overline{z}$. In fact
\begin{eqnarray}
\frac{\partial E}{\partial \overline{z}} &=&\frac{\partial
}{\partial x} (E_{x}-iE_{y})\ \frac{\partial x}{\partial
\overline{z}}+\frac{\partial }{
\partial y}(E_{x}-iE_{y})\ \frac{\partial y}{\partial \overline{z}} \\
&=&\frac{1}{2}(\partial _{x}E_{x}+\partial _{y}E_{y})\
+\frac{i}{2}(\partial _{y}E_{x}-\partial _{x}E_{y})\ .
\end{eqnarray}
Therefore, the Eqs.~(\ref{eq6}, \ref{eq7}) mean
\begin{equation}
{\it Im}\left[\frac{\partial E}{\partial\overline{z}}\right] =0\
,\qquad{\it Re}\left[\frac{\partial D}{\partial
\overline{z}}\right]=0\ . \label{integrability}
\end{equation}
The former equations can be understood as integrability conditions
for the 1-form
\begin{equation}
dw\ =\ \frac{1}{2}\ (E+D)\ dz+\frac{1}{2}\
(\overline{E}-\overline{D})\ d\overline{z}\ .
\label{complexpotential}
\end{equation}
In fact, Eqs.~(\ref{eq6}, \ref{eq7}) cancel out the exterior
derivative of the right term in Eq.~(\ref{complexpotential}). This
assures the existence of a complex potential
\begin{equation}
w(z,\overline{z})\ =\ u(x,y)+i\,v(x,y)\ .  \label{complexpotential2}
\end{equation}
By separating the real and imaginary parts of
Eq.~(\ref{complexpotential}), one obtains
\begin{equation}
du\ =\ E_{x}\ dx+E_{y}\ dy\ =\ F\ ,  \label{realpotential}
\end{equation}
\begin{equation}
dv\ =\ D_{x}\ dy-D_{y}\ dx\ =\ -\ast \mathcal{F}\ .
\label{vpotential}
\end{equation}
In the last equality we use the Hodge star operator in 2 Euclidean
dimensions:
\begin{equation}
\ast dx=-dy\, ,\qquad\ast dy=dx\, ,\qquad\ast dz=i\ dz\ .
\end{equation}
According to Eq.~(\ref{realpotential}), $(-u)$ is the usual
electrostatic potential.\footnote{As a consequence of
Eqs.~(\ref{eq7}, \ref{eq8}), $u(x,y)$ fulfills the minimal surface
equation \cite{Cou}.} Besides, the curves $v=$\textit{constant} are
field lines for $\mathbf{E}$ and $\mathbf{D}$ (they are parallel).
In fact $dv=0$ in Eq.~(\ref{vpotential}) implies that
$dy/dx=D_y/D_x$.

The Born-Infeld electrostatic problem reduces to find those complex
non-holomorphic potentials $w(z,\overline{z})$ whose exterior
derivatives adopt the form (\ref{complexpotential}), where $E$ and
$D$ are related as in Eq.~(\ref{eq8}). Contrarily, in the Coulombian
theory it is $E=D$; so the Eq.~(\ref{complexpotential}) reduces to
$dw=E\ dz$. In such case, any holomorphic function $w(z)$ provides a
Coulomb field $E=dw/dz$.

The problem of working out the Born-Infeld complex potential
$w(z,\overline{z})$ can be better tackled in terms of the inverse
function $z=z(w,\overline{w})$. \ For this, one inverts the linear
relation between $(dw,d\overline{w})$ and $(dz,d\overline{z})$;
according to Eq.~(\ref{complexpotential}), it results
\begin{equation}
dz=\frac{(\overline{E}+\overline{D})\
dw-(\overline{E}-\overline{D})\
d\overline{w}}{\overline{E}\,D+E\,\overline{D}}\ .  \label{dzeta}
\end{equation}
The relation (\ref{eq8}) between $E$ and $D$ is accomplished if both
fields are written in the following way:
\begin{equation}
E=\frac{2b}{\frac{2b}{e}+\frac{\overline{e}}{2b}}\,,\qquad
D=\frac{2b}{\frac{2b}{e}-\frac{\overline{e}}{2b}}\ ,
\label{electric}
\end{equation}
where $e(z,\overline{z})$\ is an auxiliary complex function. Notice
that $ \arg [e]=\arg [E]=\arg [D]$; so, if regarded as a vector, $e$
is colinear with $E$ and $D$. Moreover, $E=D=e$ if $b\rightarrow
\infty $, i.e. in the Coulombian limit. Replacing (\ref{electric})
in (\ref{dzeta}), it results
\begin{equation}
dz=\frac{dw}{e(w)}+\frac{\overline{e(w)}}{4b^{2}}\ d\overline{w}
\label{potencial}
\end{equation}
(cf. References \cite{ferr,Lip}). Remarkably, due to the
integrability requirement for $z(w,\overline{w})$ in
Eq.~(\ref{potencial}), the complex function $e$ depends just on $w$:
$e(z,\overline{z})=e(w(z,\overline{z}))$.\footnote{In fact, the
integrability condition $\partial (1/e)/\partial
\overline{w}=1/(4b^{2})\,\partial (\overline{e})/\partial w$ implies
$\partial e/\partial \overline{w}=0$ (proof: take absolute value).}
So, $e$ is a holomorphic function of $w$ (except, possibly, at some
singular points).

In summary, the strategy to obtain Born-Infeld electrostatic
configurations consists in: i) choose a holomorphic function $e(w)$
and integrate the Eq.~(\ref{potencial}) to obtain
$z(w,\overline{w})$; ii) solve the former relation for
$w(z,\overline{z})$ to get the complex potential $w=u(x,y)+i\,
v(x,y)$; iii) the field $E(x,y)$ can be computed by differentiating
$u(x,y)$ (see Eq.~(\ref{realpotential})) or replacing
$e(z,\overline{z})=e(w(z,\overline{z}))$ in Eq.~(\ref{electric}). If
the Born-Infeld configuration is constrained to reproduce a given
Coulombian configuration in the weak field region, then we should
use a seed function $e(w)$ that reproduces the corresponding
Coulombian relation $e_{C}(w)$ when $b\rightarrow \infty $. Of the
three steps, the second one can result unfeasible in an analytic
way. Even so, the function $z(w,\overline{w})$ of the step (i) is
useful to get the field lines. In fact, ${\it Im}[w]$ should be set
to a constant $v_{o}$ to obtain the field lines as $z=z(u,v_{o}),$
where the potential $u$ is a parameter on the field line labeled by
$v_{o}$.

As an alternative equivalent strategy, the Eq.~(\ref{potencial}) can
be rewritten as
\begin{equation}
dz=\frac{1}{e}\ \frac{dw(e)}{de}\ de+\frac{1}{4b^{2}}\
\overline{\left( e\ \frac{dw(e)}{de}\right) }\ d\overline{e}\ .
\label{potential1}
\end{equation}
where $w(e)$ is the seed, whose integration produces directly the
function $z=z(e,\overline{e})$. If this relation can be solved for
$e(z,\overline{z})$, then we replace $e(z,\overline{z})$ in
Eq.~(\ref{eq8}) to obtain the electric field $E$ as a function of
the Cartesian coordinates. Unfortunately, often this relation will
remain in the implicit form $z=z(e,\overline{e})$.

Eq.~(\ref{electric}) shows that $|\mathbf{E|}$ reaches its upper
bound limit $|\mathbf{E}|=b$\ at $|e|=2b$. Instead, $\mathbf{D}$
diverges at $|e|=2b$. Since $\mathbf{\nabla}\cdot \mathbf{D}=0$,
except at the singular points, then the flux of $\mathbf{D}$
measures the charge inside a region. This flux is
\begin{equation}
2\pi \,Q=\oint (D_{x}\ dy\ -D_{y}\ dx)=\oint dv=\oint {\it Im}[dw]\
,
\end{equation}
(the normal vector $n_{x}\,d\ell =dy$, $n_{y}\,d\ell =-dx$ is
exterior for a counterclockwise oriented path). Since
Eq.~(\ref{eq6}) implies that the circulation of the electric field
is null, then
\begin{equation}
0=\oint (E_{x}\ dx\ +E_{y}\ dy)=\oint du=\oint {\it Re}[dw]\ .
\end{equation}
Thus
\begin{equation}
2\pi \,i\ Q=\oint dw=[\Delta w]_{\Gamma}\ ,  \label{charge}
\end{equation}
where $\Gamma $ stands for the closed path in the $z$-plane. Notice
that Eq.~(\ref{charge}) is shared with Coulombian electrostatics.
However the relation between $dw$ and $dz$ is now governed by the
Eq.~(\ref{potencial}). The integral $\oint dw$ must be imaginary or
zero for a solution to be physically admissible.

\section{The monopole}
Let us exemplify the procedure with the monopolar Coulombian
potential playing the role of the holomorphic seed. In this case,
the procedure will lead to a circular symmetric Born-Infeld solution
(this solution can be straightforwardly obtained from the (real)
field equations (\ref{eq6}, \ref{eq7}); we just use it to practice
the complex calculus procedure). The Coulombian potential for the
monopole in 2 dimensions is $u_{C}=\lambda\, \log (r/r_{o})$,
$r=\sqrt{x^{2}+y^{2}}=|z|$, which is the real part of the
holomorphic complex potential $w_{C}(z)=\lambda\, {\rm Log}
(z/r_{o})$. So $e_{C}=dw_{C}/dz=\lambda /z$. Then, we will start the
procedure by choosing the Coulombian seed
\begin{equation}
w(e)=-\lambda \ {\rm Log} \left[ \frac{r_{o}\ e}{\lambda }\right]\ .
\label{monopotential}
\end{equation}
Therefore, Eq.~(\ref{potential1}) becomes
\begin{equation}
dz=-\frac{\lambda }{e^{2}}\ de-\frac{\lambda }{4b^{2}}\
d\overline{e}\ .
\end{equation}
Thus, one obtains
\begin{equation}
z=\frac{\lambda }{2b}\ \left( \frac{2b}{e}-\frac{\overline{e}}{\
2b}\right)\ . \label{zmono}
\end{equation}
From this equation and its complex conjugate, one solves the complex
field $e(z,\overline{z})$
\begin{equation}
e(z,\overline{z})=\frac{2}{1+\sqrt{\ 1+\frac{\lambda ^{2}}{b^{2}\!\
|z|^{2}}}}\ \frac{\lambda }{z}\ .
\end{equation}
To obtain the monopolar Born-Infeld electric field, we replace
$e(z,\overline{z})$ in Eq.~(\ref{electric}):
\begin{equation}
E_{x}-iE_{y}=\frac{1}{\sqrt{\ 1+\frac{\lambda ^{2}}{b^{2}\!\
|z|^{2}}}}\ \frac{\lambda }{z}=\frac{\lambda }{\sqrt{\
x^{2}+y^{2}+\frac{\lambda ^{2}}{b^{2}\!}}}\ \frac{x-i\
y}{\sqrt{x^{2}+y^{2}}}\ .  \label{monopole}
\end{equation}
Eq.~(\ref{monopole}) says that the monopolar Born-Infeld field
$E_{x}-iE_{y}$ does not diverge but behaves as $b\overline{z}/|z|$
at the origin, and recovers its Coulombian form $\lambda /z$ in the
region where $|z|\,>>\lambda /b$. On the contrary, $D$ keeps its
Coulombian form:
\begin{equation}
D_{x}-iD_{y}=\frac{\lambda }{z}=\lambda \ \frac{x-i\
y}{x^{2}+y^{2}}\ . \label{desplazamiento3}
\end{equation}
Replacing $e(w)=\lambda \ r_{o}^{-1}\exp [-w/\lambda ]$ in
Eq.~(\ref{potencial}) one obtains
\begin{equation}
z=r_{o}\ \exp [w/\lambda ]-\frac{\lambda ^{2}}{4b^{2}r_{o}}\ \exp
[-\overline{w}/\lambda]\ .
\end{equation}
Then, the Born-Infeld complex potential is
\begin{equation}
w(z,\overline{z})=\lambda \  {\rm Log} \left[ \frac{z}{2\
r_{o}}\left( 1+\sqrt{1+\frac{\lambda ^{2}}{b^{2}\ |z|^{2}}}\right)
\right]\ .
\end{equation}
To compute the charge (\ref{charge}) we surround the origin with the
counterclockwise oriented path $z=z_{o}\exp [i\,\vartheta ]$, $0\leq
\vartheta <2\,\pi $. Then
\begin{equation}
dw=\lambda \ i\ d\vartheta\ .
\end{equation}
Therefore the charge is $Q=\lambda $. The charge can also be
obtained by integrating $dw$ in the $e$-plane: when the charge is
surrounded in a counterclockwise direction, the field $e$ also
describes a circle in a counterclockwise direction. If the potential
(\ref{monopotential}) is evaluated on the path $e=e_{o}\,\exp
[i\,\vartheta ]$, $0\leq \vartheta <2\,\pi $, then the result
$dw=\lambda \ i\ d\vartheta $ is recovered.

\section{Multipoles}
\begin{figure}[t]
\centering
\includegraphics[width=15cm]{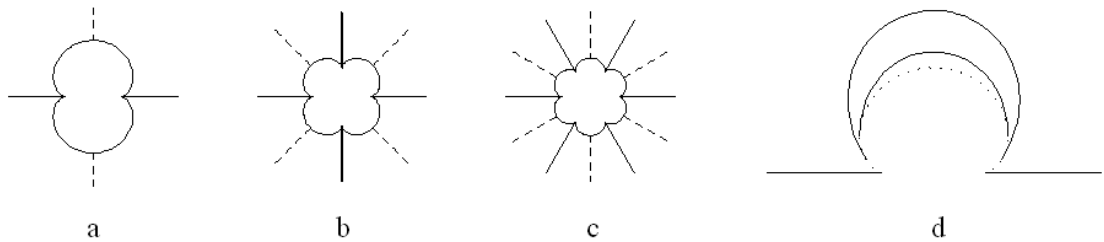}
\caption{Equipotential lines $u=0$ (dashed lines), field lines $v=0$
and epicycloids for (a) $n=1$, (b) $n=2$ and (c) $n=3$. (d) The
field lines end at the epicycloid (dotted line) (case $n=1$).}
\label{Fig1}
\end{figure}

The former example seems to confer a special value to the Coulombian
seed as a trigger of the procedure to obtain Born-Infeld solutions.
However, the direct use of the Coulombian potential as the seed not
always leads to such a satisfactory result. Let us explain this by
showing the results for the multipoles. The Coulombian potential for
the $2^{n}$-pole configuration in 2 dimensions is \ $u_{C}=-A\
r^{-n}\cos n\varphi $, $n\geqslant 1$, where $(r,\varphi )$ are
polar coordinates. \ So, the complex Coulombian potential is
$w_{C}=-A\ z^{-n}$, and the field is $e_{C}=dw_{C}/dz=nA\
z^{-(n+1)}$. Then the Coulombian seed is
\begin{equation}
w_{C}(e)=-A\left( \frac{e}{nA}\right) ^{\frac{n}{n+1}}\ .
\label{coulombian}
\end{equation}
In this case, the integration of the Eq.~(\ref{potential1}) yields
\begin{equation}
z=\left( \frac{n\ A}{e}\right) ^{\frac{1}{n+1}}-\frac{\ n^{2}A^{2}}{
4b^{2}(2n+1)}\left( \frac{\overline{e}}{nA}\right)
^{\frac{2n+1}{n+1}}\ . \label{zmulti}
\end{equation}
As an unpleasant feature of this solution, we find that the upper
bound limit $|\mathbf{E}|=b$ (i.e., $|e|=2b$) is attained not at
isolated points but at a singular closed curve surrounding the
origin (remember that $\mathbf{D}$ is still singular at the places
where $|\mathbf{E}|=b$). In fact, replacing $e=2b\ \exp
[-i(n+1)\theta ]$ in Eq.~(\ref{zmulti}) it is obtained
\begin{equation}
z(\theta )=\left( \frac{n\ A}{2b}\right) ^{\frac{1}{n+1}}\left[ \exp
[i\theta ]-\frac{\exp [i\theta (2n+1)]}{2n+1}\right]\ ,
\end{equation}
which is the parametrization of a $2n$-cusped epicycloid. Figure
\ref{Fig1}(a-c) shows the curves for $n=1,2,3$. The field lines
$v(x,y)=$\textit{constant} are obtained by integrating the
Eq.~(\ref{potencial}) for $e=e_{C}(w)=nA(-w/A)^{(n+1)/n}$:
\begin{equation}
z=\left( -\frac{w}{A}\right) ^{-\frac{1}{n}}-\frac{n^{2}\
\overline{w}^{2}}{4\ (2n+1)\ b^{2}}\ \left(
-\frac{\overline{w}}{A}\right) ^{\frac{1}{n}}\ .
\end{equation}
Thus, by setting ${\it Im}[w]$ to a constant $v_{o}$ one obtains the
field lines as $z=z(u,v_{o})$, where the potential $u$ plays the
role of the parameter on the field line labeled by $v_{o}$. Figure
\ref{Fig1}(d) shows the dipole field lines. It can be seen that the
field lines do not end at the cusps but they are tangent to the
epicycloid \cite{ferr}. The presence of a singular closed curve
where the field lines end is an unexpected feature of the solution
(\ref{zmulti}) that prevents from finding the respective inner
solution (Eq.~(\ref{eq6}) compels the inner field to be also tangent
to the epicycloid, and attain there its upper bound limit $|e|=2b$).
This trouble could be removed by choosing a different seed.
Actually, the Coulombian seed is not mandatory. We could use any
other seed recovering the Coulombian behavior when $|e|<<2b$. Thus,
it is worth asking whether a better seed could be able of reducing
the singular curve to a point. To fulfill this requirement, $dz$ in
Eq.~(\ref{potencial}) should vanish if $|e|=2b$ \cite{Pryce1}. Let
us rewrite the Eq.~(\ref{potencial}) in the form
\begin{equation}
2b\ dz=-e\ \frac{dw(e)}{de}\ d\left( \frac{2b}{e}\right) +\
\overline{\left( e\ \frac{dw(e)}{de}\right) }\ d\left(
\frac{\overline{e}}{2b}\right)\ . \label{dzmodified}
\end{equation}
Those points of the $e$-plane lying on the circle $|e|=2b$ satisfy
$(2b)/e=\overline{e}/(2b)$. So, if ${\it Im}[e\ dw/de]=0$ on the
circle $|e|=2b$, then $dz$ vanishes.\footnote{ Notice that $e\
dw/de$ is real for the monopole field.} To satisfy this reality
condition we will substitute the Coulombian seed,
\begin{equation}
e\ \frac{dw_{C}}{de}=-\frac{n\ A}{n+1}\ \left( \frac{e}{nA}\right)
^{\frac{n}{n+1}}\ ,
\end{equation}
with the improved seed
\begin{equation}
e\ \frac{dw}{de}=-\frac{n\ A}{n+1}\ \frac{\left(
\frac{2b}{nA}\right) ^{\frac{n}{n+1}}}{\left[ \left(
\frac{2b}{e}\right) ^{\frac{\alpha \ n}{n+1}}+\left(
\frac{e}{2b}\right) ^{\frac{\alpha \ n}{n+1}}\right] ^{1/\alpha}}\ .
\label{multipoleseed}
\end{equation}
For $\alpha >0$, this seed recovers the Coulombian form in the limit
$b\rightarrow \infty $. Besides it is real on the circle $|e|=2b$
because it is $(2b)/e=\overline{e}/(2b)$ ($A$ is assumed to be
positive). Let us show the behavior of the seed on the circle
$|e|=2b$ by replacing $e=2b\ \exp [-i(n+1)\theta ]$ in
Eq.~(\ref{multipoleseed}):
\begin{equation}
e\ \frac{dw}{de}\propto \frac{1}{\cos ^{1/\alpha }[\alpha
\,n\,\theta ]}\ .
\end{equation}
So the improved seed is divergent at $\alpha \,\theta _{k}=k\,\pi
/(2n)$ ($k$ is odd) and $dz$ remains indeterminate there. Therefore,
the curve where the field attain its upper bound limit $|e|=2b$
cannot be reduced to a point. The improved seed
(\ref{multipoleseed}) just substitutes the singular epicycloid by
curves (actually straight lines) where the maximal field possesses
the discretized directions $(n+1)\theta _{k}$. Let us consider this
result in the light of the simpler dipole case. For $n=1$, it is
$\alpha\,\theta=\pm\pi/2$; so, if $\alpha=1$ is chosen, then the
singular curve is reduced to a straight line where the field has
direction $\pm\pi$. This means that the $n=1$ epicycloid has been
reduced to the segment joining both cusps in Figure \ref{Fig1}(a).
In general, the choice $\alpha=1$ substitutes the singular
epicycloid for a symmetric $2n$-vertexes polygonal closed curve
whose sides coincide with the maximal field directions $(n+1)\theta
_{k}$, ($k$ is odd). Thus, the vertexes become the only sources of
field lines (point-like charges). In sum, except for the dipole
case, the inner region is not removed. However, the fact that the
curve separating the outer and inner regions now coincides with
maximal field lines creates the proper conditions to continuously
match the inner and outer solutions.

The integration of Eq.~(\ref{dzmodified}) with the seed
(\ref{multipoleseed}) yields $z(e,\overline{e})$:
\begin{eqnarray}
z=&&\left( \frac{nA}{2b}\right) ^{\frac{1}{n+1}}\ \Bigg[ \left(
\frac{2b}{e} \right) ^{\frac{1}{n+1}}F\left(
-\frac{1}{2n},1;1-\frac{1}{2n};-\left( \frac{e}{2b}\right)
^{\frac{2n}{n+1}}\right)-\nonumber\\ &&-\frac{\left(
\frac{\overline{e}}{2b} \right) ^{\frac{2n+1}{n+1}}}{2n+1}\ F\left(
1+\frac{1}{2n},1;2+\frac{1}{2n} ;-\left(
\frac{\overline{e}}{2b}\right) ^{\frac{2n}{n+1}}\right)\Bigg]\ ,
\label{zmodifiedmulti}
\end{eqnarray}
where $F(a,b,c;\xi )$ is the hypergeometric function. The expression
(\ref{zmodifiedmulti}) cannot be inverted to obtain the field
$e(z,\overline{z})$, which is left in this implicit form.
Remarkably,  $F(a,b,c;\xi)$ $=$ $1+{\cal O}(\xi)$; then, the leading
Born-Infeld correction comes from the first term in the bracket,
being of order $b^{-\frac{2n}{n+1}}$. Instead, if the Coulombian
seed were used then the Born-Infeld correction would come only from
the second term in Eq.~(\ref{potential1}), so being of order
$b^{-2}$ (see Eq.~(\ref{zmulti})). This difference is due to the
presence of $b$ in the improved seed (\ref{multipoleseed}), as a
consequence of a boundary condition ensuring the point-like
character of the charges.

The Eq.~(\ref{multipoleseed}) can be integrated to get the complex
potential:
\begin{equation}
w(e)=-A\left( \frac{e}{nA}\right) ^{\frac{n}{n+1}}\;F\left(
\frac{1}{2},1; \frac{3}{2};-\left( \frac{e}{2b}\right)
^{\frac{2n}{n+1}}\right)=-A\, \left( \frac{2b}{nA}\right)
^{\frac{n}{n+1}}\ \arctan \left[ \left( \frac{e}{2b} \right)
^{\frac{n}{n+1}}\right]\ ,  \label{BImultipotential}
\end{equation}
or
\begin{equation}
\left( \frac{e(w)}{2b}\right) ^{\frac{n}{n+1}}=\tan \left[ -\left(
\frac{nA}{2b}\right) ^{\frac{n}{n+1}}\ \frac{w}{A}\right]\ .
\label{BIfield}
\end{equation}
\begin{figure}[t]
\centering
\includegraphics[width=15cm]{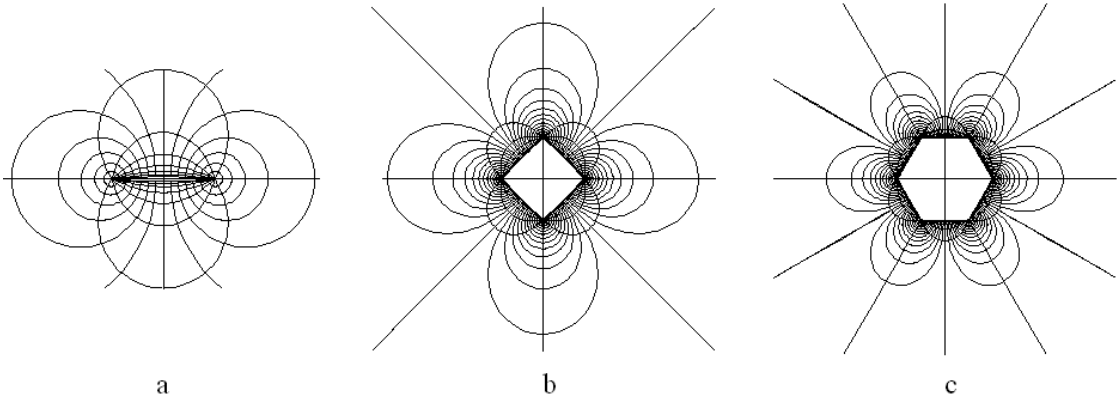}
\caption{Equipotential and field lines for Born-Infeld $2^n$-pole
configurations: (a) $n=1$, (b) $n=2$, (c) $n=3$. }\label{Fig2}
\end{figure}
By substituting this function in the Eq.~(\ref{zmodifiedmulti}) we
obtain $z=z(w,\overline{w})$. By fixing ${\it Im}[w]=v_{o}$ we
obtain the field lines $z(u,v_{o})$ as curves parametrized by the
potential $u$ and labeled by $v_{o}$. Figure \ref{Fig2} shows the
field lines for the cases $n=1,2,3$. In the case $n=1$, the maximum
field is attained at the segment joining the two opposite charges.
In the rest of the cases the maximum field lines form regular
polygons which display charges of alternate signs in their $2n$
vertexes. On the sides of these polygons the field is $e_{k}=2b\
\exp [-i(n+1)\theta _{k}]=2b\ \exp [-i(n+1)k\pi /(2n)]$, $k$ is odd
($|k|\leq 2n-1$). By replacing this field in the complex potential
(\ref{BImultipotential}), we get $v=\pm \infty $ on the polygon. The
sizes of the polygons are obtained by replacing $e=2b$ in
Eq.~(\ref{zmodifiedmulti}). The hypergeometric function
$F(a,b,c;\xi)$ is multivaluated; its principal branch has a cut on
the real axis for $1\leq \xi <\infty$. When evaluated at $e=2b$,
Eq.~(\ref{zmodifiedmulti}) gives the position of the charge lying on
the positive $x$-semiaxis:
\begin{equation}
z|_{e=2b}=\left( \frac{nA}{2b}\right) ^{\frac{1}{n+1}}\ \frac{\pi
}{2n\ \sin [\frac{\pi }{2n}]}\ .
\end{equation}
On this charge the potential $u$ reaches the bound value
\begin{equation}
u_{b}=-A\, \left( \frac{2b}{nA}\right)
^{\frac{n}{n+1}}\frac{\pi}{4}\ .
\end{equation}
At infinity the complex potential $w$ goes to zero.

\section{The multipolar inner solutions}
In order to complete the multipolar solutions $n>1$, we should fill
the interior of the polygons with a field that continuously matches
the outer field on the polygon boundary. So, we should start by
choosing an inner seed preserving the symmetry of the configuration.
Let us try the complex potential $w=-B\ z^{n}$, $B>0$, which
possesses the same symmetries that the Coulombian outer potential.
Then $e=dw/dz=-Bn\ z^{n-1}$, and so it is $w=-B(-e/nB)^{n/(n-1)}$.
Notice that the potential and the field are negative on the positive
$x$-semiaxis, as it should be expected to properly match with the
outer solution. Therefore
\begin{equation}
e\ \frac{dw}{de}=-\frac{nB}{n-1}\ \left(\exp{[i\pi]}\
\frac{e}{nB}\right) ^{\frac{n}{n-1}}\ .
\end{equation}
We will improve this expression by changing it for
\begin{equation}
e\ \frac{dw}{de}=-\frac{nB}{n-1}\  \frac{\left( \frac{2b}{nB}\right)
^{\frac{n}{n-1}}}{\left(\exp{[-i\pi]}\ \frac{2b}{e}\right)
^{\frac{n}{n-1}}+\left(\exp{[i\pi]}\ \frac{e}{2b} \right)
^{\frac{n}{n-1}}}\ . \label{innerseed}
\end{equation}
The inner seed (\ref{innerseed}) is real on the circle $e=2b\ \exp
[-i\theta (n-1)-i\pi]$, and indeterminate for $\theta _{k}=k\pi
/(2n)$ ($k$ odd). The integration of Eq.~(\ref{potential1}) can be
linked to the outer solution by changing $n\rightarrow -n$ and
$e\rightarrow \exp{[i\pi]}\, e $. Then,
\begin{eqnarray}
z=&&\left( \frac{2b}{nB}\right) ^{\frac{1}{n-1}}\Bigg[\left(
\frac{e}{2b}\, \exp{[i\pi]}\right) ^{\frac{1}{n-1}}F\left(
\frac{1}{2n},1;1+\frac{1}{2n};-\left( \frac{e}{2b}\,
\exp{[i\pi]}\right) ^{\frac{2n}{n-1}}\right)+\nonumber\\
&&+\frac{\left( \frac{\overline{e}}{2b}\, \exp{[-i\pi]} \right)
^{\frac{2n-1}{n-1}}}{2n-1}\ F\left( 1-\frac{1}{2n},1;2-\frac{1}{2n}
;-\left( \frac{\overline{e}}{2b}\, \exp{[-i\pi]}\right)
^{\frac{2n}{n-1}}\right) \Bigg]\ . \label{zmodifiedmultiinner}
\end{eqnarray}
The complex potential $w(e)$ is
\begin{equation}
w(e)=-B\, \left( \frac{2b}{nB}\right) ^{\frac{n}{n-1}}\ \arctan
\left[ \left( \frac{e}{2b}\, \exp{[i\pi]}\right)
^{\frac{n}{n-1}}\right]\ . \label{BIinnerpotential}
\end{equation}
\begin{figure}[t]
\centering
\includegraphics[width=12cm]{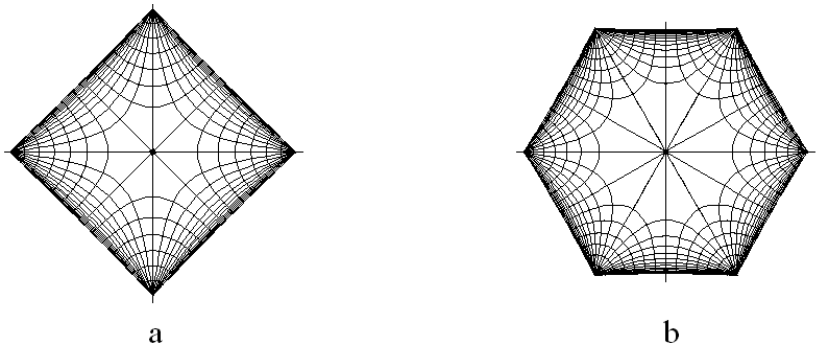}
\caption{Born-Infeld $2^n$-pole configurations (inner solution): (a)
$n=2$, (b) $n=3$. }\label{Fig3}
\end{figure}
The field lines $z(u,v_{o})$ can be obtained by replacing
$e(w(u,v_{o})) $ in the Eq.~(\ref{zmodifiedmultiinner}). Figure
\ref{Fig3} shows the inner field lines for the cases $n=2$ and
$n=3$. To properly join this inner solution with the outer solution,
we will match the positions of the charges by equalizing the points
where the field is maximum. For this purpose, it will be enough to
consider the solution on the positive $x$-semiaxis. There the field
$e$ is real; it varies from $0$ to $-2b$, when going from the center
to the vertex, and it varies from $2b$ to $0$ when going from the
vertex to infinity. It is worth noticing that the evaluation of the
bracket in Eq.~(\ref{zmodifiedmulti}) at $e=2b$ gives the same value
that the evaluation  of the bracket in
Eq.~(\ref{zmodifiedmultiinner}) at $e=-2b$. Therefore, the inner and
the outer solutions match if
\begin{equation}
\left( \frac{2b}{nB}\right) ^{\frac{1}{n-1}}=\left(
\frac{nA}{2b}\right) ^{\frac{1}{n+1}}\ \label{sizes}.
\end{equation}
This relation also guarantees the continuity of the potential along
the polygonal curve (cf. Eqs.~(\ref{BImultipotential}) and
(\ref{BIinnerpotential})). Of course, the field $E$ is continuous
too. In fact, the improved seeds (\ref{multipoleseed}) --with
$\alpha=1$-- and (\ref{innerseed}) have the ability of reducing the
singular $e=2b$ curves to polygonal curves whose sides coincide with
maximal field lines. This allows the continuity of $E$ both in
direction and magnitude, whenever the Eq.~(\ref{sizes}) assures that
the sizes of the inner and outer field structures fit each other. On
the other hand, $D$ diverges on the $e=2b$ polygonal curve. This
means that the vertexes cannot be regarded as isolated monopoles,
although they are the sources of all the field lines. Actually they
are strongly tied in a whole multipolar structure: as shown in
Section 8, the divergence of $D$ entails an infinite force on each
charge (cf. Eq.~(\ref{force4}) with the divergent result
(\ref{force5})).

\section{The dipole}
We will rework the dipole case ($n=1$). Following the
Eq.~(\ref{BIfield}), the function $e(w)$ is
\begin{equation}
e=2b\ \tan ^{2}\left[ -\frac{w}{\sqrt{2bA}}\right]. \label{edipole}
\end{equation}
We replace it in Eq.~(\ref{potencial}) to obtain
$z=z(w,\overline{w})$ as
\begin{equation}
z=-\sqrt{\frac{A}{2b}}\ \left( \cot \left[
\frac{w}{\sqrt{2bA}}\right] + \frac{w}{\sqrt{2bA}}-\tan \left[
\frac{\overline{w}}{\sqrt{2bA}}\right] +
\frac{\overline{w}}{\sqrt{2bA}}\right)\ .  \label{zmodifieddipole}
\end{equation}
Notice that $w/\sqrt{2bA}=\pm \,\pi /4$ implies that $e=2b$; then
the charges are located at
\begin{equation}
z|_{e=2b}=\pm\ \sqrt{\frac{A}{2b}}\ \ \frac{\pi}{2}\ .
\end{equation}
On the segment between the charges, the potential $u/\sqrt{2bA}$
varies in the range $[-\pi /4,\pi /4]$, while $v=\pm \infty $. On
the rest of the $x$-axis it is $v=0$. One can easily verify that the
equipotential curves $z=z(u_{o},v)$ are the circles
\begin{equation}
\left\vert \ z+\frac{u_{o}}{b}+\sqrt{\frac{A}{2b}}\ \cot \left[
\frac{2\,u_{o}}{\sqrt{2bA}}\right] \right\vert ^{2}=\frac{A}{2b}\
\cot ^{2}\left[ \frac{2\,u_{o}}{\sqrt{2bA}}\right]\ .
\end{equation}
This is the solution studied in Section XI of Ref.~\cite{Pryce1}.

The function $z=z(e,\overline{e})$ is obtained by substituting the
potential $w(e)$ in the Eq.~(\ref{zmodifieddipole}):
\begin{equation}
z=\sqrt{\frac{A}{2b}}\ \left( \left( \frac{e}{2b}\right)
^{-\frac{1}{2}}+\arctan \left[ \left( \frac{e}{2b}\right)
^{\frac{1}{2}}\right] -\overline{\left( \frac{e}{2b}\right)
^{\frac{1}{2}}}+\arctan \left[ \overline{\left( \frac{e}{2b}\right)
^{\frac{1}{2}}}\right] \right)\ . \label{zdipole}
\end{equation}
If the field $e$ does a closed path around the origin in the
$e$-plane, then $z\rightarrow -z$; so, a double turn around the
origin in the $e$-plane completes a turn around the dipole in the
$z$-plane. However, just one trip rounding the origin, but passing
$e=-2b$, corresponds to a complete trip around a charge in the
$z$-plane (passing by $e=-2b$ means crossing over the segment
between the charges). The surrounded charge is infinite, since
$-\infty <v<\infty $ (see Eq.~(\ref{charge})). This conclusion is
also valid for the other multipoles.

\section{Electrostatic force}
The force $\mathbf{P}$ on the charges inside a region is the flux of
the stress tensor on the boundary of the region. In 2 dimensions, it
is
\begin{equation}
P^{i}=-\oint_{\Gamma }\ T^{\,i\,j}\ n_{j}\ d\ell \ ,
\end{equation}
where the normal vector $n_{x}\,d\ell =dy$, $n_{y}\,d\ell =-dx$ is
exterior for a counterclockwise oriented path $\Gamma$ surrounding
the charges (the flux is zero whenever no charges are surrounded).
According to Eq.~(\ref{temunu}), it is
\begin{equation}
P\equiv P_{x}-i\,P_{y}=\frac{1}{4\pi }\oint_{\Gamma }\left[
E(D_{x}\,dy-D_{y}\,dx\,)-i\,b^{2}(dx-i\,dy)\left(
1-\sqrt{1-\frac{|E|^{2}}{b^{2}}}\right) \right]\ . \label{force4}
\end{equation}
We will use Eqs.~(\ref{complexpotential2}, \ref{vpotential}) to
replace $D_{x}\,dy-D_{y}\,dx$, and Eq.~(\ref{potencial}) to
substitute $dx-i\,dy$; thus
\begin{equation}
P=\frac{1}{4\pi }\oint \left[ E\ \frac{dw-d\overline{w}}{2\,i}
-i\,b^{2}\left( \frac{d\overline{w}}{\overline{e}}+\frac{e}{4b^{2}}
\,dw\right) \left( 1-\sqrt{1-\frac{|E|^{2}}{b^{2}}}\right) \right] \
. \label{force}
\end{equation}
We can use Eq.~(\ref{electric}) to replace $E$. In particular, it is
\begin{equation}
1-\sqrt{1-\frac{|\mathbf{E}|^{2}}{b^{2}}}=\frac{1}{2b^{2}}\
\frac{|e|^{2}}{1+\frac{|e|^{2}}{4\,b^{2}}}\ .
\end{equation}
Then, Eq.~(\ref{force}) reduces to
\begin{equation}
P=-\frac{i}{8\pi }\ \oint \ e\ dw\ .  \label{force1}
\end{equation}
To compute the force (\ref{force1}) between the dipole charges, one
can surround a charge by choosing $\Gamma$ as the closed path in the
$z$-plane formed by the $y$-axis and a semi-circle at infinity. The
field is Coulombian on the semi-circle at infinity: $|e|\sim r^{-2}$
and $|w|\sim r^{-1}$; thus, the flux at infinity vanishes. So, the
force will come from the flux on the $y$-axis, where $u=0$, and
$0<|v|<\infty $. By using the Eq.~(\ref{edipole}), the force
(\ref{force1}) is written as the integral
\begin{equation}
P=-\frac{i}{4\pi }\ \int\limits_{-\infty }^{0}\ 2b\ \tan ^{2}\left[
-\frac{i\,v}{\sqrt{2bA}}\right] \ d(i\,v)=-\frac{b}{2\pi
}\int\limits_{-\infty }^{0}\tanh ^{2}\left[
\frac{\,v}{\sqrt{2bA}}\right] \ dv\ ,\label{force5}
\end{equation}
which diverges.

\section{Electrostatic energy}
The energy density $T^{00}$ of a Born-Infeld electrostatic field is
(see Eq.~(\ref{temunu}))
\begin{equation}
T^{00}=\frac{1}{4\pi }\,\ \mathbf{E\cdot D}-\frac{b^{2}}{4\pi }\
\left( 1-\sqrt{1-\frac{|\mathbf{E}|^{2}}{b^{2}}}\right)\ .
\label{energydensity}
\end{equation}
Born and Infeld succeeded in getting a finite self-energy for the
three dimensional point-like charge because the first term in
Eq.~(\ref{energydensity}) diverges at the origin in a softer way
than in Maxwell's theory. This is the benefic effect of the regular
behavior of $\mathbf{E}$ at the origin, even though the monopolar
field $\mathbf{D}$ keeps its Coulombian form as mentioned in Section
2.\footnote{ In two dimensions it remains a logarithmic divergence
at infinity, since both $\mathbf{D}$ and $\mathbf{E}$ go to zero in
the $1/r$ Coulombian way.} This successful performance at the level
of a monopole could break down for other multipoles, because the
Coulombian divergence of $\mathbf{D}$ at the origin becomes more
dramatic. However, the solutions obtained in Sections 3 and 4 show
that Born-Infeld electrostatics spreads the multipolar sources in a
set of individual charges on a polygonal curve. So, there is a hope
that self-energies remain finite even for multipolar configurations.
In terms of $e$, the electrostatic energy density
(\ref{energydensity}) is
\begin{equation}
T^{00}=\frac{1}{4\pi }\,\left[
\frac{|e|^{2}}{1-\frac{|e|^{4}}{16\,b^{4}}}-
\frac{1}{2}\frac{|e|^{2}}{1+\frac{|e|^{2}}{4\,b^{2}}}\right] \ =\
\frac{\frac{|e|^{2}}{8\,\pi }}{1-\frac{|e|^{2}}{4b^{2}}}\ .
\label{2denergydensity}
\end{equation}
On the other hand, the volume is
\begin{equation}
dx\wedge dy=\frac{dz+d\overline{z}}{2}\ \wedge\
\frac{dz-d\overline{z}}{2\,i}\ =\ \frac{i}{2}\ dz\wedge
d\overline{z}\ .
\end{equation}
We will integrate the energy density (\ref{2denergydensity}) in the
$z$-plane to obtain the electrostatic energy. We can also change the
integration to the $w$-plane by using the Eq.~(\ref{potencial}):
\begin{equation}
dz\wedge d\overline{z} =\left(
\frac{dw}{e}+\frac{\overline{e}}{4b^{2}}\,d\overline{w}\right)
\wedge \left(
\frac{d\overline{w}}{\overline{e}}+\frac{e}{4b^{2}}\,dw\right)
=\frac{1}{|e|^{2}}\,\left( 1-\frac{|e|^{4}}{16\,b^{4}}\right)
\,dw\wedge d\overline{w}\ .
\end{equation}
Therefore
\begin{equation}
T^{00}\ dx\wedge dy=\frac{i}{16\,\pi }\ \left(
1+\frac{|e|^{2}}{4\,b^{2}} \right)\ dw\wedge d\overline{w}\ ,
\label{2denergydensity1}
\end{equation}
where $(i/2)\,dw\wedge d\overline{w}$ is the volume in the
$w$-plane.

We will work out the integration of the density
(\ref{2denergydensity1}) for the dipole configuration. Using the
Eq.(\ref{edipole}), it is
\begin{equation}
T^{00}\ dx\wedge dy=\frac{i}{16\,\pi }\ \left( 1+\left\vert \tan
\left[ - \frac{w}{\sqrt{2bA}}\right] \right\vert ^{4}\right)
dw\wedge d\overline{w}\ ,
\end{equation}
where $(i/2)\,dw\wedge d\overline{w}=du\,dv$. In the semi-plane
$x<0$, it is $0<u<\sqrt{2bA}\pi /4$, $-\infty <v<\infty $. Then, the
dipole electrostatic energy is
\begin{equation}
U =\int \ T^{00}\ dx\wedge dy =\frac{Ab}{2\,\pi }\ \int_{-\infty
}^{\infty }dv\ \int_{0}^{\pi /4}du\ \left( 1+\left\vert \tan \left[
u+i\ v\right] \right\vert ^{4}\right)\ .
\end{equation}
The integral on the variable $v$ is divergent.

\section{Two opposite isolated charges}
Let us now consider the Coulombian ingredients for the field of two
equal but opposite charges $\lambda$, $-\lambda$ separated by a
distance $d$:
\begin{equation}
w_{C}=\lambda\  {\rm Log} \left[
\frac{z-\frac{d}{2}}{z+\frac{d}{2}}\right]\ .
\end{equation}
Then
\begin{equation}
e_{C}=\frac{dw_{C}}{dz}=\frac{\lambda \ d}{z^{2}-\left(
\frac{d}{2}\right) ^{2}}\ ,
\end{equation}
\begin{equation}
e\ \frac{dw_{C}}{de}=-\frac{\lambda \
\sqrt{e}}{\sqrt{e+\frac{4\lambda }{d}}} =-\frac{\sqrt{\lambda \
d}}{2\ \sqrt{\frac{d}{4\lambda }+\frac{1}{e}}}\
.\label{Coulombianseparated}
\end{equation}
This last expression should be substituted by an improved seed
accomplishing the reality condition. Additionally, one should
require that the dipole field be recovered for $d\rightarrow 0$,
$\lambda \rightarrow \infty $ (but $\lambda d$ remaining a
constant). Even so, the answer seems not to be unique (see, however,
Ref.~\cite{Pryce2} and the comments included in footnote 6 and
Section 11). We choose
\begin{equation}
e\ \frac{dw}{de}=-\ \frac{\sqrt{\frac{b\, \lambda \,
d}{2}}}{\sqrt{\frac{bd}{2\lambda }+\left(
\sqrt{\frac{2b}{e}}+\sqrt{\frac{e}{2b}}\right) ^{2}}}\ ,
\end{equation}
which has the right Coulombian limit, it goes to the Born-Infeld
dipole for $d\rightarrow 0$ and $\lambda\rightarrow\infty$ (but
$A=\lambda d$), and it is real on the circle $|e|=2b$. By expanding
the binomial one gets
\begin{equation}
e\ \frac{dw}{de}=-\frac{\sqrt{\frac{b\, \lambda \,
d}{2}}}{\sqrt{\frac{bd}{2\lambda }+2+\frac{2b}{e}+\frac{e}{2b}}}\ ,
\label{separatedcharges}
\end{equation}
which is the case studied in Section IX (example 3) of
Ref.~\cite{Pryce1}. For $d\rightarrow\infty$, one recovers the
isotropic monopolar expression $e\, dw/de=-\lambda$ at every point
where $e\neq 0$ (see Section 4). Instead, for finite values of $d$,
$e\, dw/de$ is not isotropic in the $e$-plane even at the circle
$e=2b\, \exp[i \vartheta]$ --i.e., at the charges--, where the
radicand becomes $bd/(2\lambda)+2\,(1+\cos\vartheta)$. This is a
characteristic feature of Born-Infeld solutions. On the contrary,
the Coulombian field diverges at the charges; thus, the Coulombian
expression (\ref{Coulombianseparated}) becomes monopolar-like at the
charges. As it will be explained in the Conclusions, this
non-isotropic behavior leads to the single-valuedness of the
Born-Infeld field $e(z, \overline{z})$.

\begin{figure}[t]
\centering
\includegraphics[width=10.5cm]{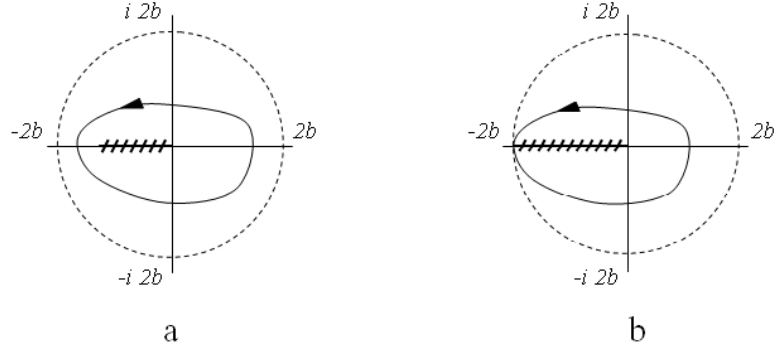}
\caption{Branch cuts of the multivalued function $e\, dw/de$: (a)
two equal but opposite charges; (b) dipole. }\label{Fig4}
\end{figure}

Let us review the benefits of passing from the Coulombian seed
(\ref{Coulombianseparated}) to the improved seed
(\ref{separatedcharges}) from a different point of view. One of the
consequences is the splitting of the singularity at $e=-4\lambda/d$
in the Coulombian seed into two singularities $e_{1,2}$ on the
negative real axis of the $e$-plane:
\begin{equation}
\frac{e_{1,2}}{2b}=-a\pm\sqrt{a^2-1}\, ,\qquad a\equiv
1+\frac{bd}{4\lambda}\, .\label{roots}
\end{equation}
Notice that it is $|e_1|<2b$ and $|e_2|>2b$: $e_1$ and $e_2$ are
respectively inside and outside the circle $|e|=2b$. Actually $e_1$
is the value of $e$ at the center of the configuration. In fact, the
symmetry of the configuration implies that $e$ is real and negative
only on the $y$-axis and the segment between the charges. Besides,
on the $y$-axis it is $u=0$ (like in the Coulombian case), while on
the segment between the charges it is $v=$\textsl{constant}. So, for
$e<0$, $dw$ passes from being imaginary to becoming real at $z=0$.
This change of behavior happens when the radicand in the
Eq.~(\ref{separatedcharges}) changes sign, i.e. at $e=e_1$. So,
$e_1$ is the value of $e$ at $z=0$ (if $b\rightarrow \infty$, then
$e_1$ goes to the Coulombian value $-4\lambda /d$). The improved
function (\ref{separatedcharges}),
\begin{equation}
e\ \frac{dw}{de}=-b\, \sqrt{\lambda \, d}\
\frac{\sqrt{e}}{\sqrt{(e-e_1)\, (e-e_2)}}\ ,
\label{separatedcharges1}
\end{equation}
is multivalued; it has a branch cut inside the circle $|e|=2b$
between $e_1$ and $0$ (it has also a branch cut outside the circle
between $e_2$ and $\infty$). This cut means that there are two
different ways of surrounding $e=0$ inside the circle. If the path
crosses the cut (i.e., if the path is close to $e=0$), then the
function (\ref{separatedcharges}) will return to its initial value
after two turns. This behavior is typical of a dipolar structure:
far from the charges, where the field is near to zero, a path in the
$z$-plane closes after two complete turns of the field; this feature
is reflected by the function (\ref{separatedcharges1}) which is
directly related to the position via the
Eq.~(\ref{potential1}).\footnote{If a closed path in the $z$-plane
surrounds a $2^n$-polar structure, then the initial field $e_o$
changes to $e_o\, \exp[\pm i\, 4n\pi]$ once the closed path is
completed. Conversely, $2n$ turns are needed in the $e$-plane to
come back to the initial position.} However, if the sources are
constrained to be just isolated charges, it should be possible to
surround an individual charge and find a structure similar to a
monopole (no branch cuts in such case). This is the reason why the
branch cut cannot reach the circle $|e|=2b$: since $2b$ is the upper
bound for $|e|$, which is attained at the charge positions, then
there must exist closed paths in the $e$-plane near (but inside) the
circle $|e|=2b$ that do not cross any branch cut of the seed $e\,
dw/de$. On the contrary, the dipole displays a branch cut that
extends from $-2b$ to $0$; in such case, the only way of surrounding
an individual charge is passing the $e=-2b$ point (i.e., crossing
the dipole singular segment). In both cases --two opposite separated
charges and dipole-- the multivalued function $e\, dw/de$ has two
Riemann sheets in the domain $|e|\leq 2b$, which correspond to each
surroundable charge. Figure \ref{Fig4}(a,b) shows the $e$-plane for
two opposite separated charges and the dipole, respectively; it also
includes a path surrounding an individual charge. We conclude that
any physically meaningful Born-Infeld configuration generated by
isolated charges must allow for closed paths near the circle
$|e|=2b$ which do not cross any branch cut of the seed $e\, dw/de$.
Nevertheless, some branch cuts inside the circle are needed to open
a Riemann sheet for each individual charge. Since the functions of
the form $e^m\, dw/de$ are well behaved in a ring including the
circle $|e|=2b$, then Cauchy-Goursat theorem states that their
integrals on closed paths that goes near the circle $|e|=2b$
(without crossing branch cuts) are independent of the path. So, this
kind of integrals can be performed directly on the
circle.\footnote{It is worth mentioning that all these remarkable
properties would be spoiled if other types of seed were chosen. The
Born-Infeld seed (\ref{separatedcharges}) reproduces inside the
circle the structure of singularities and branch cuts that the
Coulombian seed has in the whole $e$-plane.} As an application, let
us compute the individual charges in the Born-Infeld field of two
opposite separated charges. We use Eq.~(\ref{charge})
\begin{equation}
2\pi \,i\ Q=\oint_{\Gamma }dw=\oint_{\Gamma }\frac{dw}{de}\ de\ ,
\end{equation}
where $\Gamma $ is a counterclockwise path surrounding the charge in
the $z$-plane. In the $e$-plane, it corresponds to a closed path
near the circle; so we will integrate on the circle: $e=2b \exp[i
\vartheta]$. According to the Eq.~(\ref{separatedcharges}), it is
\begin{equation}
2\pi \,i\ Q =\pm \sqrt{\frac{b\lambda\ d}{2}}\int_{0}^{2\pi}
\frac{i\ d\vartheta}{\sqrt{2(a+\cos{\vartheta})}}\ .
\end{equation}
So, the value of $Q$ is
\begin{equation}
Q=\pm \frac{\lambda }{\pi }\sqrt{a-1}\ \left[ \frac{K\left(
\sqrt{\frac{-2}{a-1}}\right) }{\sqrt{a-1}}+\frac{K\left(
\sqrt{\frac{2}{a+1}}\right) }{\sqrt{a+1}}\right]\ ,
\end{equation}
where $K(k)$ is the complete elliptic integral of the first kind
(here we follow the notation of Refs.~\cite{Abra,Grad}).
$|Q|$ ranges from $0$, for $a\rightarrow 1_+$, to $\lambda$ for
$a\rightarrow\infty$. Therefore, the individual charges are smaller
than those suggested by the far (Coulombian) field.

We can also compute the force (\ref{force1}) on an individual charge
by integrating on the circle $e=2b\, \exp[i\, \vartheta]$. Then, we
use Eq.~(\ref{separatedcharges}) to write the force as
\begin{equation}
P=\mp\frac{i}{8\pi}\oint\, e\, \frac{dw}{de}\ de=\pm\frac{i\,
\lambda }{8\pi}\sqrt{2(a-1)}\int_{0}^{2\pi}\frac{2b\, i\, exp[i\,
\vartheta]}{\sqrt{2(a+\cos[\vartheta])}}\ d\vartheta\ .
\end{equation}
Therefore,
\begin{equation}
P=\mp\frac{\lambda\, b\, \sqrt{a-1}}{4\, \pi}\ \left[-\frac{2\,
K\left( \sqrt{\frac{-2}{a-1}}\right) }{\sqrt{a-1}}+\frac{2\, K\left(
\sqrt{\frac{2}{a+1}}\right) }{\sqrt{a+1}}+\frac{\pi\,
F\left(\frac{1}{2},\frac{3}{2},2; \frac{-2}{a-1}\right)
}{\sqrt{a-1}}-\frac{\pi\, F\left(\frac{1}{2},\frac{3}{2},2;
\frac{2}{a+1}\right) }{\sqrt{a+1}}\right].\label{force0}
\end{equation}
This expression goes to $\pm \lambda^2/(2d)$ when
$a\rightarrow\infty$ (Coulombian limit). But it vanishes when $a$
goes to $1_+$ (i.e., when the charges approach each other). This
result is consistent with the vanishing of the charges when they go
together. The force reaches its maximum value at $a=1.15746$, i.e.
at $d=0.63\, \lambda/b$. It is always lower than the Coulombian
force; at the leading order in $b^{-1}$ it is
\begin{equation}
P=\pm\, \frac{\lambda^2}{2\, d}\, \left(1-\frac{6\, \lambda}{b\,
d}\right)+{\cal O}(b^{-2})\ .\label{force2}
\end{equation}
However, except for the Coulombian limit, $d$ does not coincide with
the real distance between the charges. The distance should be
computed by integrating the Eq.~(\ref{potential1}). The field $e$ on
the $x$-axis is real and varies from $e_1$ at $x=0$ to $-2b$ at the
position $x_L$ of the left charge. Then
\begin{equation}
x_L=\int_{e_1}^{-2b}\left[ \frac{1}{e}\,
\frac{dw}{de}+\frac{e}{4b^2}\, \frac{dw}{de} \right]\, de
=-\frac{\lambda}{2b}\, \sqrt{2(a-1)} \,
\int_{\frac{e_1}{2b}}^{-1}\frac{1+\varepsilon^{-2}}{\sqrt{2a
+\varepsilon^{-1}+\varepsilon}}\ d\varepsilon\, ,
\end{equation}
where $e_1/(2b)=-a+\sqrt{a^2-1}$ (see Eq.~(\ref{roots})). This is an
involved integral. Nevertheless, it can be verified that
$x_L\rightarrow 0$ if $d\rightarrow 0$, and $x_L\rightarrow -d/2$ in
the Coulombian limit.

\section{Conclusions}
We have displayed a method to obtain Born-Infeld electrostatic
solutions in 2 dimensions, which is condensed in the paragraph after
the Eq.~(\ref{potencial}). This procedure is a cleaner version of
the one developed by Pryce in Section II of Ref. \cite{Pryce1}. The
method starts from a holomorphic (except at some isolated points)
seed $w(e)$, where $w$ is the complex potential and $e$ is a complex
variable linked to the electric field $E$, to then obtain a
non-holomorphic function $z(e,\overline{e})$ connecting the field
with the Cartesian coordinates $z=x+i\, y$. Although any seed $w(e)$
could be employed, one should prescribe that $w(e)$ goes to the
Coulombian potential $w_C(e)$ for $b\rightarrow\infty$, in order to
reobtain the Coulombian field in the weak field region. Moreover,
$e\, dw/de$ has to be real on the circle $|e|=2b$ in order that the
field sources correspond just to isolated points. The way of
achieving this reality condition conferred interesting symmetries to
the seed $w(e)$ and the resulting function $z(e,\overline{e})$. In
fact we have chosen seeds $w(e)$ such that $e\,dw/de$ does not
change under the transformation $e/(2b)\rightarrow 2b/e$ (see
Eqs.~(\ref{multipoleseed}), (\ref{innerseed}) and
(\ref{separatedcharges})). On the circle $|e|=2b $, $e/(2b)$ and
$2b/e$ are complex conjugate. So $e\,dw/de$ is real on the circle,
which is the requirement to get isolated singularities. Therefore,
\begin{equation}
e\ \frac{dw}{de}=\frac{e}{2b}\,
\frac{dw(\frac{e}{2b})}{d(\frac{e}{2b})}=\frac{2b}{e}\
\frac{dw(\frac{2b}{e})}{d(\frac{2b}{e})}=\frac{1}{e}\
\frac{dw(\frac{2b}{e})}{d(\frac{1}{e})}=-e\
\frac{dw(\frac{2b}{e})}{de}\, ,
\end{equation}
that it can be integrated to obtain
\begin{equation}
w\left( \frac{e}{2b}\right) =-w\left( \frac{2b}{e}\right) +
constant. \label{property}
\end{equation}
The complex potentials (\ref{BImultipotential}) and
(\ref{BIinnerpotential}) effectively possess this property since
\begin{equation}
\arctan [\xi ]=-\arctan \left[\frac{1}{\xi }\right]+\frac{\pi }{2}\
,\hspace{0.3in}\xi \in\mathbb{C}.
\end{equation}
On the other hand, the Coulombian monopolar potential
(\ref{monopotential}) has already the property (\ref{property}); so,
it does not need any improvement. In Eq.~(\ref{dzmodified}), the
symmetry in question implies that the function $z(e,\overline{e})$
has the form
\begin{equation}
z(e,\overline{e})=f\left( \frac{2b}{e}\right) -f\left(
\frac{\overline{e}}{2b}\right) + constant.  \label{property2}
\end{equation}
where
\begin{equation}
f^{\prime}(\xi)=f^{\prime}\left(\frac{1}{\xi }\right).
\label{property3}
\end{equation}

Properties (\ref{property2}, \ref{property3}) are evident in the
monopolar solution (\ref{zmono}). For the solutions
(\ref{zmodifiedmulti}) and (\ref{zmodifiedmultiinner}), the
properties are verified by means of the identity
\begin{equation}
\xi^{\frac{1}{n+1}}F\left(-\frac{1}{2n},1;1-\frac{1}{2n};-\frac{1}{\xi}
\right)=\frac{\xi^{\frac{2n+1}{n+1}}}{2n+1}\, F\left(
1+\frac{1}{2n},1;2+\frac{1}{2n};-\xi^{\frac{2n}{n+1}}\right)
+\frac{\pi}{2n\,\sin[\frac{\pi }{2n}]}\ ,
\end{equation}
where $-\pi <\arg [\xi ]<\pi $.

It was also explained in Section 10 that the chosen seeds caused
that those closed paths in the $e$-plane going near the circle
$|e|=2b$ do not cross the branch cuts of functions $f^{\prime}$. As
a consequence, those integrals surrounding individual charges can be
performed directly on the circle (Cauchy-Goursat theorem). The
functions $z(e, \overline{e})$ possessing all these characteristics
guarantee the single-valuedness of the field $e(z, \overline{z})$.
In fact, let us surround a charge and use the properties
(\ref{property2}, \ref{property3}):
\begin{equation}
\oint dz=\oint f^{\prime}\left(\frac{1}{\xi}\right)\,
d\left(\frac{1}{\xi}\right)-\oint f^{\prime}(\overline{\xi})\,
d\overline{\xi}\ ,
\end{equation}
where $\xi=e/(2b)$. Both integrals in the right side are equal. In
fact, since the integrands are well behaved near the circle, then
the closed path can be deformed into the circle $\xi=\exp[i\,
\vartheta]$, where the integrals become manifestly equal. Therefore
it is $\oint dz=0$, i.e., closed paths in the $e$-plane are also
closed paths in the $z$-plane, which means that the field is
single-valued.

Finally, we have shown that the force between two equal but opposite
charges $\pm \lambda$ reaches the maximum value at $d=0.63\,
\lambda/b$, but then it decreases up to vanish when the charges
approach each other ($d$, $\lambda$ are not the actual distance and
charge, but their magnitudes inferred from the far (Coulombian)
field). In the weak field region, the interaction force departs from
its Coulombian partner at the first order in $b^{-1}$ (see
Eq.~(\ref{force2})). As remarked in Section 5, the corrections of
order lower than $2$ do not come from the Eq.~(\ref{potential1}),
but originate in boundary conditions to guarantee the point-like
character of the sources.

\bigskip

\acknowledgments The author is grateful to Mauricio Leston for
helpful discussions.

\end{document}